\documentstyle[11pt,newpasp,twoside,epsf]{article}
\def \msol {\rm{M}$_\odot$} 
\def \msols {\rm{M}$_\odot$~yr$^{-1}$} 
\def\kms{km~$\rm{s}^{-1}$}

\def\ie{{\it ie.~}}
\def\ea{{\it et al.~}}

\def\edcomment#1{\iffalse\marginpar{\raggedright\sl#1\/}\else\relax\fi}
\reversemarginpar

\begin{document}
\title{A Paradigm Lost: New Theories for Aspherical PNe}
\author{Adam Frank}
\affil{Department of Physics and Astronomy, 
University of Rochester, Rochester NY 14627}

\begin{abstract}
Theoretical Models for the shaping of PNe are reviewed in light of new
high resolution images. The new data indicate the
purely hydrodynamic interacting stellar winds model can not recover
the full variety of shapes and kinematics.  New models, some speculative,
others more firmly grounded are discussed.  In particular, 
accretion disks and magnetic fields are identified as two of the most
promising avenues of future research.  Outstanding issues such as jet
formation by PNe disks and dynamo activity in P-AGB stars remain to
studied.  Finally, new simulations of the Egg Nebula are presented
as an example of a ``paleontological'' study designed to recover the
history of an individual object.
\end{abstract}

\section{Introduction}

In the last decade Planetary Nebulae (PNe) have gone from mysterious to
mundane and back again.  Ten years ago the variety of PNe shapes was
seen as puzzling and lacked a unified physical theory.  Five years later it
appeared that the needed theoretical interpretation had been found in the
Generalized Interacting Stellar Winds (GISW) model.  The
wealth of new, high resolution space and ground based observations of
PNe has made the GISW paradigm seem less universal than once hoped
for and these objects once again present us with significant
puzzles and theoretical challenges.  It is remarkable that such a
common phenomena as the end stages of low and intermediate  stars
should so consistently surprise and outwit us.   In this paper my aim is
to review the state of aspherical PNe theory.  I am particularly
interested in where the new observations show the limits of the GISW
model and where, in my opinion, the next generation of theoretical
models is likely to emerge.

\section{Observations Outpace Theory: Limits of the GISW Model}
Researchers in this community are, by now, well aware of the main
contours of the GISW model. I repeat them for completeness (for
more details see Kwok \ea 1978,  Kahn \&
West 1986, Balick 1987, Icke 1988 and also Frank 1999 for a review).
The defining paradigm for explaining aspherical PNe posits a single
star evolving from the AGB to a white dwarf.  As the star evolves so
does its wind.  A slow ($10$ \kms), dense ($10^{-4}$ \msol ) wind
expelled during the AGB is followed by a fast ($1000$ \kms), tenuous
($10^{-7}$ \msol ) wind driven off the contracting proto-white dwarf
during the PNe phase.  The fast wind expands into the slow wind
shocking and compressing it into a dense shell.  The ``Generalized''
part of this interacting winds scenario occurs when the AGB wind
assumes a toroidal geometry with higher densities in the equator than
at the poles. Inertial gradients then confine the expanding shock
leading to a wide range of morphologies.  Numerical models
(Soker \& Livio 1988, Mellema \ea 1991, Icke \ea 1992, Frank \& Mellema
1994, Mellema \& Frank 1995, Dwarkadas \ea 1996), have shown this
paradigm can embrace a wide variety of nebular morphologies including
(and this point needs to be stressed) {\em highly collimated jets}
(Icke \ea 1992, Frank \& Mellema 1997, Mellema \& Frank 1998, Borkowski
\ea 1997). In spite of the success of these models, high resolution
observations, primarily from the HST, have revealed new features of PNe
morphology which {\it appear} difficult to recover with the classic
GISW model.  Below I list what I consider to be the most
vexing issues raised by these observations.

{\bf Point-symmetry:} This morphological class was first identified by
Schwarz \ea 1992.  It is most striking in PNe where jets or strings
of knots are visible. There are however many bipolar (and even some
elliptical) PNe which also show point symmetry in their brightness
distributions.  It has been suggested that point symmetry occurs due to
precession of a collimated jet (Lopez 1997).
If this is the case it is difficult to imagine that a large-scale
out-flowing gaseous torus (needed for GISW) can provide a stiff precessing
nozzle for the flow.  Inhomogeneities in the torus would tend to smooth
out on sound crossing times of $\tau = L/c \approx 10^{17} cm / 10^6
cm ~s^{-1} \approx 3 \times 10^3 y$.  This is on order of, or less than, 
many of the inferred precesession periods.

{\bf Jets and Ansae:} While the GISW model {\it can} produce narrow
jets it usually requires a large-scale ``fat'' torus.  It is not clear
if such structures exist in PNe (this point requires further
study).  The FLIERS/Ansae described by Balick \ea 1997 are more
problematic in that they have, so far, resisted most attempts at
theoretical explanation (see however Mellema \ea 1998) .

{\bf Multi-polar/Episodic Outflows:} A number of PNe show multiple,
nested bipolar outflows (Guerrero \& Manchado 1998, Welch \ea 1999).
In some cases these multiple bipolar shells have common axii of
symmetry, while in others the axii are misaligned (multi-polar).  Such objects
present problems both for the  GISW model and the ``classical'' view of
post-AGB evolution.  When the nebulae is multi-polar it is usually
point-symmetric and presents the same challenges described above.  In some
cases (Kj PNe 8, Lopez \ea 1997) it appears as if the development of
dense torus may occur more than once with separate axii in each case.
Such a phenomena is difficult to reconcile with either a classic single-star 
model or appeals to binary scenarios.  The most likely explanation
for any multiple bubbles whether uni- or multi-polar is an episodic
wind.  A fast wind occurring in outbursts or varying periodically is
not part of the standard model for post-AGB evolution
(though a nova-like recurrence is possible in binary systems).  These
nebulae may originate from born-again PNe however the time-scales for the
bubbles ($<10^4$ y) do not appear well matched with He shell
inter-pulse timescales.

{\bf Post-AGB (P-AGB)/Proto-PNe (PPNe) Bipolar Outflows:} One of the
most startling results of the last five years is the recognition that
fast ($\ge 100$ \kms) bipolar outflows can occur in the PPNe or
even the Post-AGB stage.  Objects like CRL 2688 and the Red Rectangle raise
the question of how high-velocity collimated flows occur when the
star is still in a cool giant or even supergiant stage (CRL 2688 has an
F Supergiant spectral type: Sahai \ea 1998; Young \ea 1992).  The
origin of the wind in this early stage and the mechanisms which produce
its collimation are critical questions because it appears that much of the
shaping of PNe may occur before the ``mature'' PNe phase when the star
has become hot enough to produce a strong ionizing flux (Sahai \&
Trauger 1998). 

\section{New Physics}

In spite of its successes, it appears that the GISW model can not
embrace the full range of behaviors observed in PPNe and PNe.  In this
section I briefly review (or suggest) some alternative scenarios for PNe
evolution. In all cases the scenarios focus on the source of the outflow,
either the nascent Central Star of a Planetary Nebulae (CSPNe) or an
accretion disk surrounding the star.

\subsection{Common Envelope Evolutionary Truncation (CEET)} The ``story''
of PNe evolution has long included the possibility of a common
envelope phase (Iben \& Livio 1993, Soker 1998).  When a giant star
swallows a companion the pair's common envelope can be rapidly
ejected (most likely in the orbital plane) leaving either a merged core
or a short period binary.  This process need not occur at the tip of
the AGB when the giant's stellar core has already evolved to a
CSPNe configuration.  Common envelope ejection can occur at any
point in the evolution along the AGB (or RGB) branch. The initiation of
the CE evolution depends on the size of the star {\it and} the orbital
separation of the binary.  Thus the envelope of the star can be torn off
the core exposing it before it is in a stable CSPNe configuration.

If the core is not yet ``prepared'' to be a CSPNe then we might expect
instabilities that could lead to rapid and violent ejection of material
remaining in the atmosphere.  This could provide a source
of explosive energy release.  This is an attractive idea because it may explain
the fully non-asymmetric structures such as the elephant trunks
seen in some PPNe and very young PNe (Trammel, these proceedings).
For an exposed core of mass $M_c$ with a remnant envelope of mass $M_e$ the time-scale
for thermal instabilities can be crudely represented as the
Kelvin-Helmholtz timescale (Kippenhahn \& Weigert 1989)
For typical CSPNe this is $\tau_{KH} < 10^3$ y, a short fraction of the AGB
evolutionary timescale.  Thus these instabilities could occur rapidly
allowing them to escape direct detection.  Note also that if these
stars have a strong dynamo acting then the loss of the envelope could
expose the kG fields at the base of the convective zone on short timescales 
such that magnetic instabilities could act as the source of
explosive energy release.

Finally, consider the possibility that the secondary does not spiral all
the way in to merge with the core but leaves instead an extremely close
binary (with separation $a$).  In the period just after the CE ejection
the two stars will orbit rapidly in a environment still rich with
circumbinary gas (Sandquist \ea 1998).  Spiral shocks driven by the secondary's
orbital motion will heat the gas to temperatures proportional to the
$V_k^2$ where $V_k$ is the Keplerian speed of the secondary.  If the
cooling time $t_c$ for the gas is greater or equivalent to its sound
crossing time $t_x$ then the pressure gradients will set the gas in
motion.  It is possible therefore that the close binary will produce a
kind of {\it egg-beater} effect driving a wind from the source at
speeds
\begin{equation}
V_w = \zeta V_k = \zeta \sqrt{G M_c/a}
\label{eggbv}
\end{equation}
In the equation above $\zeta \le 1$ and would depend on the ratio of
$t_c/t_x$.

While this CEET scenario is highly speculative it may
provide routes for either explosive energy release or winds in the
PPNe stage.  In lieu of other mechanisms for driving relatively fast
winds from cool PPN stars (Simis, Dominik \& Icke, these proceedings)
this feature makes the CEET a scenario worth further exploration.

\subsection{Accretion Disk Winds}

The possibility  that accretion disks play a
role in PNe formation was first suggested by Morris (1987).  More
recently Soker \& Livio (1994) and Reyes-Ruiz \& Lopez (these
proceedings) have explored the formation of disks in binary PNe
systems in more detail.  In these works there has been a tacit
assumption that disk = outflows.  Obviously there is gap in the theory
and it remains unclear if or how accretion disks
in PNe can create collimated outflows that match observations.  I now
review two classes of disk wind models that may be applicable for PNe.

{\bf Magneto-centrifugal Launching:} The potential for accretion disks
to create strong, collimated winds has been explored in some detail
by the both the YSO and AGN communities (Ouyed \& Pudritz 1997, Shu
\ea 1994).  The most popular models rely on the presence of magnetic
fields embedded in the disk (\ie the foot points of the field are
tied to the disk via surface currents). The field co-rotates with the
disk. If field lines are bent at an appropriate angle to the disk axis
($\theta > 30^o$) energy can be extracted from rotation and matter
loaded on the field lines is flung outward ``like a bead on a wire''.
This mechanism has been shown effective in both analytical and
numerical  studies.  While it is clear that the mechanism can produce
winds on the order of the escape speed at the launch point, the geometry
of the wind that is generated is still uncertain.  Some studies
indicate that a narrow jet of hypersonic plasma will form almost
immediately above the disk/star system (Ouyed \& Pudritz 1997).
Other researchers (Shu \ea 1994) find the collimation process to
be slow leading to so-called ``wide-angle'' winds with cylindrical
density stratification (the densest parts of the flow lie along the
axis giving the appearance of a jet).

Magneto-centrifugal launching has many attractive features for PNe.
For instance the presence of narrow jets in the midst of wider
bipolar flows might be naturally explained by wide-angle wind models.
The issue which must be addressed is, can such flows be established
in PNe disk systems? YSO and AGN disks have typical size scales
of hundreds of $AU$.  A PNe disk could have a size scale of order
the Roche Lobe. Can the appropriate physics be obtained in
these smaller disks?  For example, can magnetic fields of the right magnitude
and geometry be generated in these disks?  Finally, can the flows
from the disk produce the observed morphology and kinematics of the
nebulae? Recent MHD jet propagation studies using magneto-centrifugal
launching models as input show promising results in terms of generating
diverse flow characteristics (Frank \ea 1999) however such research
remains in its infancy.  

It is, at least, possible to estimate the terminal speed of a
magneto-centrifugal disk wind as 
\begin{equation} 
V_\infty \approx \Omega(r) r_A(r) 
\label{rotspeed} 
\end{equation} 
where $\Omega(r)$ is
the rotation rate in the disk at radius $r$ and the $r_A(r)$ is the
Alf\'ven radius of the flow launched at that disk radius.  Typically $r_A$
is of the order of a few times $r$ (Pelletier \& Pudritz 1992).  Thus assuming
a keplerian disk with a characteristic size of order the stellar
radius for a P-AGB star ($r \approx R_\odot$), the equation above
yields $V_\infty \approx 100 - 1000 km/s$ which is of the order
observed in PNe jets. Thus the application of MHD disk-wind models
to PNe could be a promising field for future work (Frank \ea 1999).

{\bf Radiation-driven Disk Winds:} Even without magnetic fields
accretion disks can generate outflows.  Angular momentum must be
dissipated in order to allow material in the disk to spiral inward.
Thermal energy created in the dissipational process can heat the disks'
surface layers to high enough temperature for line driving to become
effective.  A wind is then driven off the surface of the disk in the
same manner as one is produced by a hot star.

In a series of numerical models Proga and collaborators (Proga, Stone
\& Drew 1998) have explicated the properties of radiative disk winds.
They find the winds emerge with a conical geometry with large half-opening
angles of $\theta > 45^o$.  It is noteworthy that the flow pattern in the
winds can be unsteady.  In general the maximum mass loss
rates in these models tends to be low ($\dot{M} < 10^7$ \msols)
and are associated with high wind velocities ($v_w > 1000$ \kms).
Thus these winds would likely produce fairly wide lobed energy
conserving bipolar outflows.  Application of these models to the
short lived accretion disks which would likely occur in PNe have not
yet been attempted and further study in this area may prove fruitful.

\subsection{Magnetized Wind Bubbles} One of the most promising new
theoretical models invokes a toroidal magnetic field embedded in a
normal radiation driven stellar wind.  This so-called {\it Magnetized
Wind Bubble (MWB)} model was first proposed by Chevalier \& Luo (1994)
and has been studied numerically by Rozyczka \& Franco (1996) and
Garcia Segura \ea (1999).  In these models the field at the star is
dipolar but assumes a toroidal topology due to rapid stellar rotation.
When the wind passes through the inner shock, hoop stresses associated
with the toroidal field dominate over isotropic gas pressure forces
and material is drawn towards the axis producing a collimated flow.
This mechanism has been shown capable of producing a wide variety of
outflow morphologies including well collimated jets.  When precession
of the magnetic axis is included in fully 3-D simulations, the MWB
model is capable of recovering point-symmetric morphologies as well
(Garcia-Sequra 1997).  The capacity for the magnetic field to act as
a long lever arm imposing coherent structure across large distances
makes these models particularly attractive.

The potential difficulties involved in application of the MWB models
include the presence of field reversals at the ``equatorial current sheet''
(which need not be restricted to the equator) where reconnection could
produce strong dissipation of the magnetic field (Soker 1998, Frank
1999).  A more serious difficulty involves the rather extreme input
parameters required for the hoop stresses to become effective.  The
critical parameter in the MWB model is the ratio of magnetic
to kinetic energy in the wind, $\sigma$.  In terms of parameters at the stellar
surface
\begin{equation} 
\sigma = {B^2 \over 4\pi \rho_w V_w^2} = {B_*^2 R_*^2
\over \dot{M}_w V_w} \left( {V_{rot} \over V_w}\right)^2 
\label{sigeq}
\end{equation}
where $V_{rot}$ is the rotational velocity of the star.  The MWB model is
only effective when $\sigma > .01$.  It has been noted that this value
is what obtains in the Sun.  While such an identification may seem
initially seem promising for the model one must recall that the
$\dot{M}_{PN}/\dot{M}_{\odot} > 10^7$!  Thus the additional factor of ten
million or more must be made up by some combination of field strength,
rotational velocity or stellar size.  Unfortunately these are usually
anti-correlated. Given that $\dot{M}$ is fairly well established it appears
that one needs either very strong fields or very high
rotation rates.

The situation becomes more difficult when one considers that without
significant angular momentum transport in the star, mass losing AGB
stars should spin-down during their evolution.  Consider the simplest
case of a constant density star rotating as a solid body.  One can show
that given a main sequence rotation rate, mass and radius of
$\Omega_{ms},M_{ms}$ and $R_{ms}$ respectively, the post-AGB rotation
rate will be
\begin{equation}
\Omega_{P} = \Omega_{ms} \left({R_{ms} \over R_{P}}\right)^2 
\left( {M_{P} \over M_{ms}}\right)^{2/3}
\label{myeq}
\end{equation}
where $P$ denotes Post-AGB quantities.  Note that since $M_{P} <
M_{ms}$ and $R_{P} > R_{ms}$ we will always have $\Omega_{P} <
\Omega_{ms}$.

It has been argued that effective mixing between the core and the
envelope during the AGB stage can produce high surface rotational
velocities (Garcia-sequra \ea 1999).  While this may be possible it's
effectiveness would tend to diminish dynamo processes which may be needed to
create the magnetic field.

\subsection{Dynamos} If magnetic fields play a role in post-AGB and PNe
evolution then one must address the source of the field.  While it is
possible that dynamically significant fields may be preserved as fossil
relics from the main sequence, it is more likely that the fields may
be generated via dynamo processes.  The standard $\alpha-\omega$ mean
field dynamo model used to explain most astrophysical magnetic fields
(stellar, galactic etc.) relies on a combination of convection and
differential rotation.  The rotation stretches the field lines while
convection turns toroidal components into a poloidal (dipole-like)
field.  The effectiveness of a dynamo can be expressed in terms of the
dynamo number $N_D$ (Thomas, Markiel \& Van Horn 1995),
\begin{equation}
N_D \propto \Delta \Omega {r_c \over L}
\label{dynnum}
\end{equation}
where $\Delta \Omega$ is the differential rotation which occurs over a
scale $L$ in the midst of a convection zone of size $r_c$.  Effective
dynamos in AGB, post-AGB or CSPNe require $N_D > 1$.  The equation
above shows the difficulty of using a rapidly rotating star whose
angular momentum has been well mixed as the source of strong magnetic
fields. Unless the field is a fossil of the previous stages, dynamo
generated fields require strong differential rotation.

There remains much work to be done in the application of dynamo theory
to AGB stars (Pascoli 1997).  The growing implication that MHD is
a necessary part of the nebular dynamics is likely to require that
such work be done.

\section{Nebular Paleontology}

There are many examples of PNe studies in which attempts have been made
to directly link simulations with data via synthetic observations.
These have usually involved calculation of various optical forbidden
and permitted line intensity maps.  As this approach matures it should
become possible to carry out stellar wind paleontology studies in
which the history of an individual object is reconstructed based on
its morphology, kinematics, ionization and chemical structure.

For such studies to be successful they require objects that have
been very well characterized observationally ($\eta$ Car, SN1987A;
Collins et al 1998).  The Egg Nebula is becoming one such object.
Figure 1 presents the results of a paleontological study of the Egg
nebula (Delamarter \ea 1999) in which simulations that included $H_2$
chemistry and excitation as well as post-processed scattered light
image production were carried out.  After more than 50 simulations
we found that the GISW model could {\it not} recover the observed
features of the Egg.  Further simulations showed that the best fit to
$H_2$ and scattered light intensity maps required a torus ejected at
about the same time as a fully collimated jet.  The torus and jet were
distinct, non-interacting dynamical features.  Based on the requirement
that $H_2$ not become dissociated the results allowed for reasonably
unique determination of the mass loss history of the central star.
We note in Fig 1, however, that these models could not recover the
unusual tuning-fork pattern seen in the scattered light images of
the Egg.  The images did produce a good match to other PPN such as
IRAS 17150-3224.

\acknowledgements Support for this work was provided at the 
University of Rochester by NSF grant AST-9702484 and the 
Laboratory for Laser Energetics.

\clearpage
\begin{figure}
\plotone{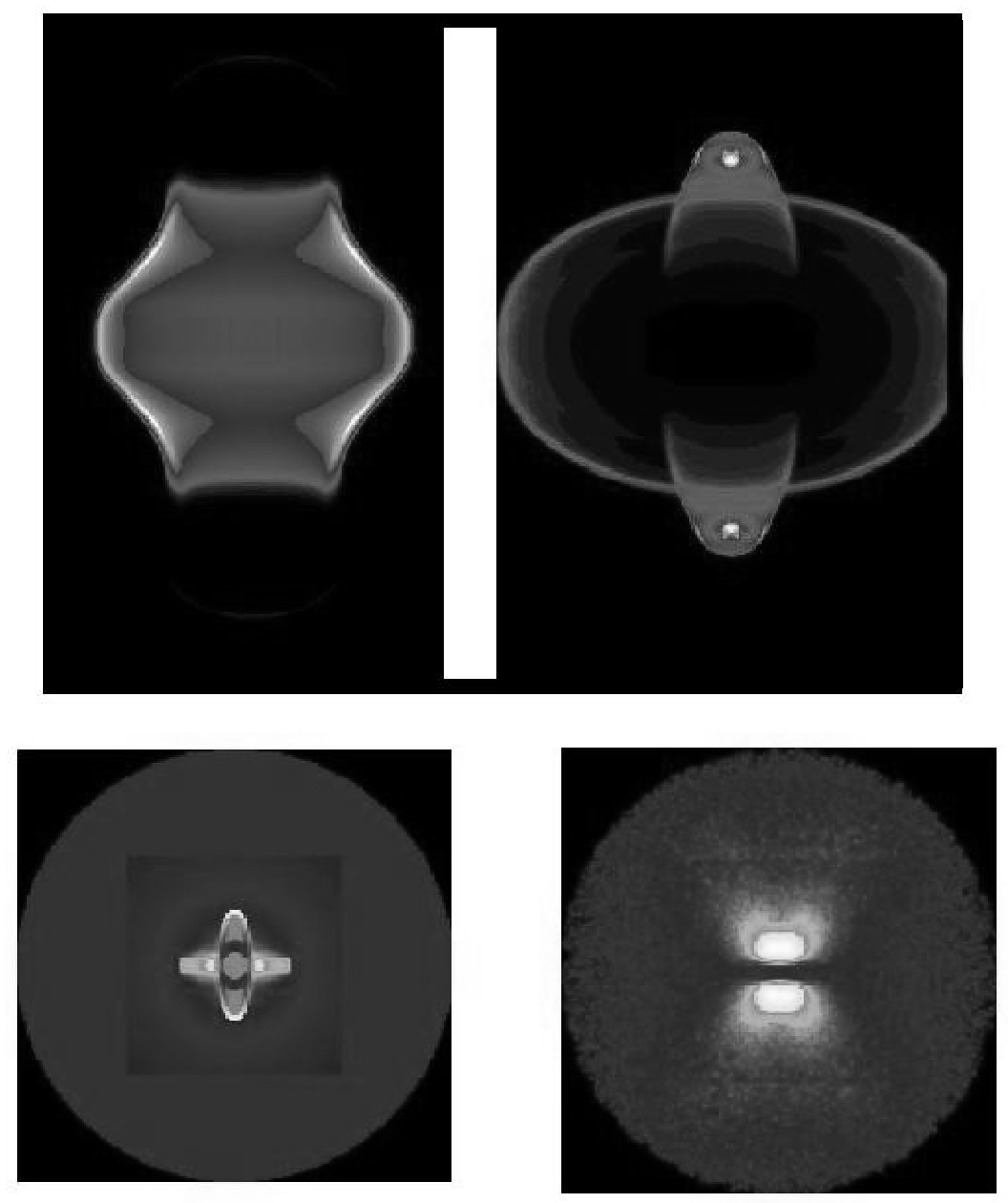}
\caption[]{Results of hydrodynamic simulations of the Egg Nebulae.
Bottom Left: Log density for a torus + jet model.  A torus of high
density gas ejected during the P-AGB phase expands into a spherically
symmetric AGB wind.  The torus is followed by a well collimated
jet propagating along the symmetry axis.  The square contours are
an artifact of an expanded grid.  Bottom Right: Scattered light
image taken from this simulation.  Upper Right: $H_2$ image of a torus +
jet model.  Note how shock emission at the edge of the torus defines
a ring.  Upper Left: $H_2$ image of a GISW model.  Note that the model fails
to capture the salient properties of the Egg.}
\end{figure}

\end{document}